\documentclass[american,prl, reprint, unsortedaddress]{revtex4-1}
\usepackage{lmodern}
\usepackage[T1]{fontenc}
\usepackage[latin9]{inputenc}
\usepackage{amsmath}
\usepackage{graphicx}
\usepackage{babel}
\begin{document}
\title{Efficient Langevin dynamics for ``noisy'' forces}
\author{Eitam Arnon}
\affiliation{Fritz Haber Research Center for Molecular Dynamics, Institute of Chemistry,
The Hebrew University of Jerusalem, Jerusalem 91904, Israel}
\author{Eran Rabani}
\affiliation{Department of Chemistry, University of California and Materials Science
Division, Lawrence Berkeley National Laboratory, Berkeley, California
94720, U.S.A.; The Raymond and Beverly Sackler Center for Computational
Molecular and Materials Sciences, Tel Aviv University, Tel Aviv, Israel
69978}
\email{eran.rabani@berkeley.edu}

\author{Daniel Neuhauser}
\affiliation{Department of Chemistry and Biochemistry, University of California
at Los Angeles, CA-90095 USA}
\email{dxn@chem.ucla.edu}

\author{Roi Baer}
\affiliation{Fritz Haber Research Center for Molecular Dynamics, Institute of Chemistry,
The Hebrew University of Jerusalem, Jerusalem 91904, Israel}
\email{roi.baer@huji.ac.il}

\selectlanguage{american}%
\begin{abstract}
Efficient Boltzmann-sampling using first-principles methods is challenging
for extended systems due to the steep scaling of electronic structure
methods with the system size. Stochastic approaches provide a gentler
system-size dependency at the cost of introducing ``noisy'' forces, which serve to limit the efficiency of the sampling. In
the first-order Langevin dynamics (FOLD), efficient sampling is achievable
by combining a well-chosen preconditioning matrix S with a time-step-bias-mitigating
propagator (Mazzola et al., Phys. Rev. Lett., 118, 015703 (2017)).
However, when forces are noisy, S is set equal to the force-covariance
matrix, a procedure which severely limits the efficiency and the stability
of the sampling. Here, we develop a new, general, optimal, and stable
sampling approach for FOLD under noisy forces. We apply it for silicon
nanocrystals treated with stochastic density functional theory and
show efficiency improvements by an order-of-magnitude.
\end{abstract}
\maketitle
Prediction of the equilibrium properties of extended systems using
atomistic models often requires sampling from the Boltzmann distribution
a series of configurations \citep{frenkel2002understanding,allen1987computer,rapaport2004theart,skeel2017comparing,gao2017transport,schlick2010molecular}.
Most common sampling methods implicitly assume that either the potential
energy surface \citep{ceperley2003metropolis} or the forces on the
nuclei \citep{marx2000abinitio,tuckerman2010statistical,ceriotti2009langevin,dai2009largescale,bussi2007accurate}
are accessible, either through deterministic ab-initio methods such
as density functional theory (DFT) or other quantum chemistry methods
(for small-medium sized systems) \citep{marx2009abinitio} or through
empirical force-fields. For extended systems, \emph{ab initio} methods
often rely on stochastic techniques such as Quantum Monte Carlo (QMC)
\citep{ceperley1999thepenalty,luo2014abinitio,mazzola2017accelerating,krajewski2006linearscaling}
or stochastic DFT (sDFT) \citep{baer2013selfaveraging,arnon2017equilibrium,neuhauser2016stochastic,cytter2018stochastic,chen2019overlapped,fabian2019stochastic}.
For example, in sDFT the forces are calculated using a relatively
small number of \emph{stochastic orbitals }instead of using the full
set of deterministic Kohn-Sham eigenstates. Therefore, the forces
calculated within sDFT are noisy with fluctuating values. Such noisy
forces can also occur with partially-converged self-consistent field
approaches to deterministic DFT \citep{kuhne2007efficient,martinez2015thermostating}.

Langevin dynamics (LD) often serves to generate a series of thermally-distributed
nuclear configurations, based on the calculated forces on the nuclei.
The balance between accuracy, which favors small time-steps, and efficiency,
which requires large time-steps (to reduce the correlations between
consecutive configurations in the series), determines the overall
complexity and accuracy of this class of approaches. A common form
of Langevin dynamics is the so-called second-order LD (SOLD) \citep{attaccalite2008stableliquid,arnon2017equilibrium,krajewski2006linearscaling,tassone1994acceleration,luo2014abinitio,bennett1975masstensor},
in which the Newton equation of motion includes a friction term and
a noisy force obeying the fluctuation-dissipation relation. An alternative
is the first-order Langevin dynamics (FOLD) \citep{risken1989thefokkerplanck,becca2017quantum,luo2014abinitio},
which is conceptually simpler than SOLD because it does not have inertia
and therefore only nuclear configurations are Boltzmann-sampled. FOLD
is amenable to the introduction of a \emph{preconditioning matrix,
}which, by proper choice, dramatically increases the configurational
sampling efficiency without affecting the accuracy \citep{risken1989thefokkerplanck}.
Unfortunately, when the forces are noisy, this preconditioning matrix
must be set equal to the force covariance matrix \citep{mazzola2017accelerating}
and, thus, cannot be used for obtaining optimal sampling efficiency.
Therefore, it seems that noisy forces, used in conjunction with FOLD,
are inherently less efficient than deterministic ones. An additional
complication appears as numerical instabilities due to the singular
nature of the force covariance matrix. 

In this letter, we develop an approach that enables the use of noisy
forces within FOLD lifting the constraints on the preconditioning
matrix. Furthermore, we demonstrate the approach for silicon NCs within
sDFT and show an order of magnitude increase in sampling efficiency
compared to state of the art methods for noisy forces. The solution
lies in \emph{adding }random noise which combines with preconditioning
matrix to complement the noise in the force coming from the stochastic
electronic structure method. 

In its simplest form, the time-discretized first-order Langevin dynamics
produces a set of $M$ configurations $\boldsymbol{R}_{\tau}\equiv\left(R_{\tau}^{1},\dots,R_{\tau}^{3N}\right)^{\dagger}$,
$\tau=1,\dots,M$ for a $N$ nuclei system, through a random walk
described by \citep{mazzola2017accelerating} : 
\begin{align}
\boldsymbol{R}_{\tau+1} & =\boldsymbol{R}_{\tau}+\sqrt{2k_{B}T\Delta_{t}}\boldsymbol{\zeta}_{\tau}+\Delta_{t}S^{-1}\boldsymbol{f}\left(\boldsymbol{R}_{\tau}\right),\label{eq:FOLD}
\end{align}
where $\boldsymbol{f}\left(\boldsymbol{R}\right)\equiv\left(f^{1}\left(\boldsymbol{R}\right),\dots,f^{3N}\left(\boldsymbol{R}\right)\right)^{\dagger}=-\boldsymbol{\nabla}V\left(\boldsymbol{R}\right)$
is the force acting on the nuclear degrees of freedom $\boldsymbol{R}$,
$\Delta_{t}$ is a unit-less time-step parameter and $S$ is an arbitrary
$3N\times3N$ symmetric positive-definite matrix. The random vector
$\boldsymbol{\zeta}_{\tau}=\left(\zeta_{\tau}^{1},\dots,\zeta_{\tau}^{3N}\right)^{\dagger}$,
with which thermal fluctuations are introduced, is distributed such
that $\left\langle \boldsymbol{\zeta}^{\tau}\right\rangle =0$ and
\begin{equation}
\left\langle \boldsymbol{\zeta}_{\tau}\boldsymbol{\zeta}_{\tau'}^{\dagger}\right\rangle =S^{-1}\delta_{\tau\tau'}.\label{eq:FOLD-zz}
\end{equation}
For \emph{any} choice of the preconditioning matrix $S$, the generated
trajectory of $M$ samples the Boltzmann distribution at temperature
$T$ in the $\Delta_{t}\to0$ and $M\to\infty$ limits \citep{becca2017quantum}.
For finite values of $M$ and $\Delta_{t}$, the configurations can
then be used to produce estimates of the thermal average of quantities
$A$: $\left\langle A\right\rangle _{T}\approx\left\langle A_{M}\right\rangle \equiv\left\langle \frac{1}{M}\sum_{\tau=1}^{M}A\left(\boldsymbol{R}_{\tau}\right)\right\rangle $.
One would expect that the variance of $A_{M}$ is $\sigma_{A,T,M}^{2}=\frac{\sigma_{A,T}^{2}}{M}$,
where $\sigma_{A,T}^{2}$ is the thermal variance in $A$ at temperature
$T$. However, since configurations $R_{\tau}$ and $R_{\tau+\tau'}$
correlate, the actual variance is much larger: $\sigma_{A,T,M}^{2}=\frac{\sigma_{A,T}^{2}}{M}\tau_{c}$
where $\tau_{c}$ is the number of correlated time-steps. The smaller
$\tau_{c}$, the more efficient is the Langevin dynamics for sampling. 

Consider now the efficiency of the method in the $T\to0$ limit for
the $3N$-dimensional harmonic oscillator $V\left(\boldsymbol{R}\right)=\frac{1}{2}\boldsymbol{R}^{\dagger}H\boldsymbol{R},$
where $H$ is the Hessian matrix ($H_{ij}=\frac{\partial^{2}V\left(\boldsymbol{R}\right)}{\partial R^{i}\partial R^{j}}$).
In this limit, the trajectory generated by Eq.~(\ref{eq:FOLD}) with
$\boldsymbol{f}\left(\boldsymbol{R}\right)=-H\boldsymbol{R}$ is given
by $\boldsymbol{R}_{\tau}=\left(1-\Delta_{t}U\right)^{\tau}\boldsymbol{R}_{0}$,
where where $U=S^{-1}H$ and $\tau=0,1,2,\dots$ enumerates the time
steps. The correlation between displacements after many time-steps
decays as $e^{-\Delta_{t}u_{\text{min}}\tau}$, where $u_{min}>0$
is the smallest eigenvalue of $U$, so $\tau_{c}\approx\left(\Delta_{t}u_{\text{min}}\right)^{-1}$.
Furthermore, the trajectory $R_{\tau}$ remains stable as long as
$u_{\text{max}}\Delta_{t}<2$, where $u_{max}$ is the largest eigenvalue
of $U$. Thus $\tau_{c}$ is limited from below by 
\begin{equation}
\tau_{c}>\frac{1}{2}\frac{u_{\text{max}}}{u_{\text{min}}}\equiv\frac{1}{2}\text{cond}\left(U\right).\label{eq:NC}
\end{equation}
It is now evident how preconditioning is important. Without it (say
$S=I$) we find $\tau_{c}>\frac{1}{2}\text{cond}\left(U\right)$ which
in typical problems can easily exceed $10^{3}$, making the random
walk very inefficient. Optimal preconditioning involves taking $S=H$,
enabling $\tau_{c}$ to be as low as $1$. However, in this case one
would have $\tau_{c}\approx\Delta_{t}^{-1}$, and since $\Delta_{t}$
has to be kept small to avoid bias, $\tau_{c}$ is often quite large
\emph{even under preconditioning}. This is where a method that reduce
the time-step bias, thus allowing $\Delta_{t}$ to grow is required.
Such a random walk was proposed in Ref.~\citenum{becca2017quantum},
based upon the exact solution for a harmonic potential. It involves
the following process:
\begin{equation}
\boldsymbol{R}_{\tau+1}=\boldsymbol{R}_{\tau}+\sqrt{2k_{B}T\Delta_{2}}\boldsymbol{\zeta}_{\tau}+\Delta_{1}S^{-1}\boldsymbol{f}\left(\boldsymbol{R}_{\tau}\right),\label{eq:RB-FOLD}
\end{equation}
employing two time-steps:

\begin{align}
\Delta_{n} & =\frac{1}{n}\left(1-e^{-n\Delta_{t}}\right),\,\,n=1,2\label{eq: DeltaA}
\end{align}
and was shown to lead to significantly lower time-step biases. We
refer to this type of random walk as ``reduced-bias FOLD'' (RB-FOLD). 

What happens when the forces are random? Can we still use RB-FOLD
and have efficient sampling? The random forces $\boldsymbol{\phi}\left(\boldsymbol{R}_{\tau}\right)=\boldsymbol{f}\left(\boldsymbol{R}_{\tau}\right)+\boldsymbol{\eta}_{\tau}$
coming from sDFT or QMC will give the deterministic force $\boldsymbol{f}\left(\boldsymbol{R}_{\tau}\right)=\left\langle \boldsymbol{\phi}\left(\boldsymbol{R}_{\tau}\right)\right\rangle $
on the average but will also involve random inseparable fluctuations
$\boldsymbol{\eta}_{\tau}=\left(\eta_{\tau}^{1},\dots,\eta_{\tau}^{3N}\right)^{\dagger}$.
Simply plugging the random force $\boldsymbol{\phi}\left(\boldsymbol{R}\right)$
into the FOLD equation will give the wrong effective dynamics $\boldsymbol{R}_{\tau+1}=\boldsymbol{R}_{\tau}+\sqrt{2k_{B}T\Delta_{t}}\left(\boldsymbol{\zeta}_{\tau}+\sqrt{\frac{\Delta_{t}}{2k_{B}T}}S^{-1}\boldsymbol{\eta}_{\tau}\right)+\Delta_{t}S^{-1}\text{\ensuremath{\boldsymbol{f}}}\left(\boldsymbol{R}_{\tau}\right)$
since the noise fluctuations $\boldsymbol{\eta}_{\tau}$ clearly cause
additional heating, violating the fluctuation-dissipation relation.
Hence, whenever one replaces $\boldsymbol{f}\left(\boldsymbol{R}\right)$
by $\boldsymbol{\phi}\left(\boldsymbol{R}\right)$ in Eq.~(\ref{eq:FOLD})
one also needs to replace $\boldsymbol{\zeta}$ of Eq.~(\ref{eq:FOLD-zz})
by a ``smaller'' fluctuation $\tilde{\boldsymbol{\zeta}}$, so the
FOLD is now 
\begin{equation}
\boldsymbol{R}_{\tau+1}=\boldsymbol{R}_{\tau}+\sqrt{2k_{B}T\Delta_{t}}\tilde{\boldsymbol{\zeta}}_{\tau}+\Delta_{t}S^{-1}\text{\ensuremath{\boldsymbol{\phi}}}\left(\boldsymbol{R}_{\tau}\right)\label{eq:FOLD-STOCH}
\end{equation}
where: 
\begin{align}
\left\langle \tilde{\boldsymbol{\zeta}}_{\tau}\tilde{\boldsymbol{\zeta}}_{\tau'}^{\dagger}\right\rangle  & =\left[S^{-1}-\frac{\Delta_{t}}{2k_{B}T}S^{-1}\text{cov}\boldsymbol{\phi}\left(\boldsymbol{R}_{\tau}\right)S^{-1}\right]\delta_{\tau,\tau'}.\label{eq:FOLD-zz-STOCH}
\end{align}
Here, $\text{cov}\boldsymbol{\phi}\left(\boldsymbol{R}_{\tau}\right)=\left\langle \boldsymbol{\eta}_{\tau}\boldsymbol{\eta}_{\tau}^{\dagger}\right\rangle $
is the force covariance matrix and it is proportional to $\frac{1}{I}$,
where $I$ is the number of stochastic iterations in the electronic
structure calculation. Note however, that the term on the right-hand
side must be positive-definite, a condition that can be achieved by
sufficient reduction of either the time step $\Delta_{t}$ or the
random force covariance. In both cases this requires additional computational
work. In Ref.~\citenum{mazzola2017accelerating} the specific choice
$S=\alpha\times\text{cov}\boldsymbol{\phi}\left(\boldsymbol{R}_{\tau}\right)$
(where $\alpha$ is a properly chosen constant) was made, which had
the appeal that $\left\langle \tilde{\boldsymbol{\zeta}}_{\tau}\tilde{\boldsymbol{\zeta}}_{\tau}^{\dagger}\right\rangle $
, like $\left\langle \boldsymbol{\zeta}_{\tau}\boldsymbol{\zeta}_{\tau}^{\dagger}\right\rangle $
of Eq.~(\ref{eq:FOLD-zz}) was proportional to $S^{-1}$. But this
choice has the folloing shortcomings: (a) $S$ is now time-dependent
and requires special treatment in the equation of motion \citep{mazzola2017accelerating};
(b) it straddles $S$, leaving no room for using it as a preconditioning
matrix for optimizing the efficiency and (c) it assumes implicitly
that the covariance matrix is invertible, which is not always the
case \footnote{See supplementary material for further discussion of this topic.}.
In light of these limitations, we advocate leaving $S$ in its original
form as an \emph{optimal preconditioning matrix} (e.g. $S\approx H$)
and using Eq.~(\ref{eq:FOLD}) with $\boldsymbol{\phi\left(\boldsymbol{R}_{\tau}\right)}$
replacing $\boldsymbol{f}\left(\boldsymbol{R}_{\tau}\right)$ and
with $\tilde{\boldsymbol{\zeta}}$ of Eq.~(\ref{eq:FOLD-zz-STOCH})
replacing $\boldsymbol{\zeta}$ of Eq.~(\ref{eq:FOLD-zz}). We refer
to this method as ``noisy-FOLD'', since it is an extension of the
FOLD method to noisy forces. A similar treatment in the case of the
random force counterpart of RB-FOLD (Eq.~\ref{eq:RB-FOLD}), to which
we refer henceforth as ``noisy-RB-FOLD'', leads to the following
FOLD 
\begin{equation}
\boldsymbol{R}_{\tau+1}=\boldsymbol{R}_{\tau}+\sqrt{2k_{B}T\Delta_{2}}\hat{\boldsymbol{\zeta}}_{\tau}+\Delta_{1}S^{-1}\boldsymbol{\phi}\left(\boldsymbol{R}_{\tau}\right),\label{eq:RB-FOLD-STOCH}
\end{equation}
where:

\begin{equation}
\left\langle \hat{\boldsymbol{\zeta}}_{\tau}\hat{\boldsymbol{\zeta}}_{\tau'}^{\dagger}\right\rangle =\left[\left(1-\frac{\Delta_{1}^{2}}{2k_{B}T\Delta_{2}}S^{-1}\text{cov}\boldsymbol{\phi}\left(\boldsymbol{R}_{\tau}\right)\right)S^{-1}\right]\delta_{\tau\tau'}.\label{eq:RB-FOLD-zz-STOCH}
\end{equation}
These two equations form the main result of this letter, since this
noisy-RB-FOLD preserves much of the flexibility in choosing the matrix
$S$ as in the RB-FOLD solution while allowing for stochastic forces.
As noted above for noisy FOLD, here too, the right-hand side of Eq.~(\ref{eq:RB-FOLD-zz-STOCH})
must be positive-definite. To enforce this condition, additional numerical
work is required, either by decreasing either the time-step or the
force covariance. The first measure, decreasing the time-step, increases
the sample correlations, so additional time-steps are needed as a
compensation. The second measure, reducing $\text{cov}\boldsymbol{\phi},$
calls for a step up in the number of stochastic electronic-structure
iterations. 

\begin{figure}[b]
\includegraphics[width=0.95\columnwidth]{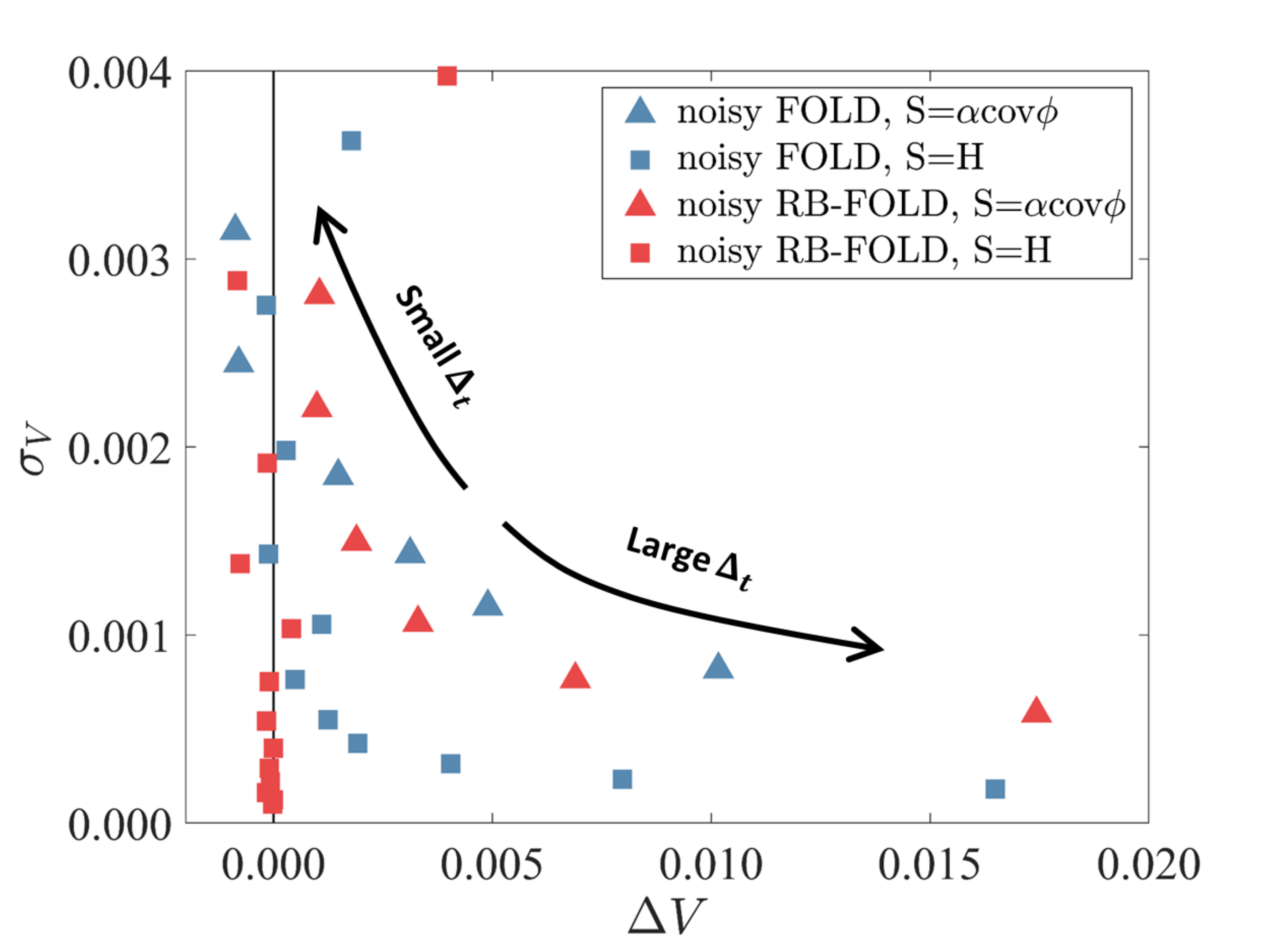}\caption{\label{fig:Harmonic oscillator}The bias ($\Delta V=\left\langle V\right\rangle -\frac{3}{2}k_{B}T$,
where $k_{B}T=0.1$) and the fluctuation $\sigma_{V}$ in the average
potential energy estimate $\left\langle V\right\rangle $ (determined
using binning analysis \citep{becca2017quantum}) for noisy- FOLD
and RB-FOLD calculations on a 3D Harmonic oscillator with a random
force $\boldsymbol{\phi}$, $\left\langle \boldsymbol{\phi}\right\rangle =H\boldsymbol{R}$
and $\text{cov}\boldsymbol{\phi}=0.02I$, where the Hessian $H$ is
diagonal with values of $0.1$, $1$ and $10$. We show results for
$S=\alpha\text{cov\ensuremath{\boldsymbol{\phi}}}$ (with $\alpha=1$,
triangles) and $S=H$ (squares). The blue symbols correspond to noisy-FOLD
(Eq.~(\ref{eq:FOLD-STOCH}),(\ref{eq:FOLD-zz-STOCH})) while the
red symbols to noisy-RB-FOLD (Eqs.~$\eqref{eq:RB-FOLD-STOCH}$-(\ref{eq:RB-FOLD-zz-STOCH})).
The points are differentiated by time a step parameter $\Delta_{t}$
(not specified). Results are calculated using trajectories of $5\times10^{7}$
steps.}
\end{figure}

We use the Harmonic potential discussed above to demonstrate the theory
in Fig.~\ref{fig:Harmonic oscillator}. We plot the fluctuation $\sigma_{V}$
and the bias $\Delta V$ for various sampling procedures within FOLD,
comparing the non-optimal preconditioning choice, $S=\alpha\text{cov}\boldsymbol{\phi}$
(with $\alpha=1$ in the units of the Harmonic oscillator, triangles)
discussed in Ref.~\citenum{mazzola2017accelerating} and the optimal
preconditioning $S=H$ (squares) advocated here. It is evident from
the figure, that whether one uses noisy-FOLD (blue symbols, Eqs.~(\ref{eq:FOLD-STOCH})-(\ref{eq:FOLD-zz-STOCH}))
or noisy-RB-FOLD (red symbols, Eqs.~(\ref{eq:RB-FOLD-STOCH})-(\ref{eq:RB-FOLD-zz-STOCH}))
the bias $\Delta V$ can be reduced only by decreasing the time steps
$\Delta_{t}$. However the above analysis of $\tau_{c}$ showed as
$\Delta_{t}$ decreases, the fluctuation $\sigma_{V}$ grows! Under
optimal preconditioning $S=H$ (squares) we see that noisy-FOLD (blue
symbols, Eqs.~(\ref{eq:FOLD-STOCH})-(\ref{eq:FOLD-zz-STOCH})) biases
are reduced yet the error control is still unsatisfactory since any
attempt to reduce the bias further (by decreasing $\Delta_{t}$) increases
once again the fluctuation $\sigma_{V}$. This problem does not arise
for noisy-RB-FOLD results (red squares, Eq.~(\ref{eq:RB-FOLD-STOCH})-(\ref{eq:RB-FOLD-zz-STOCH}))
where $\Delta_{t}$ can grow to lower $\sigma_{V}$ without a bias
penalty. Note that to within small fluctuations the same results seen
here for noisy forces also appear for deterministic ones (obtained
by taking $\phi=f$ and $\text{cov}\phi=0$, not shown here). 

\begin{figure*}[t]
\noindent\begin{minipage}[t]{1\columnwidth}%
\includegraphics[width=0.95\columnwidth]{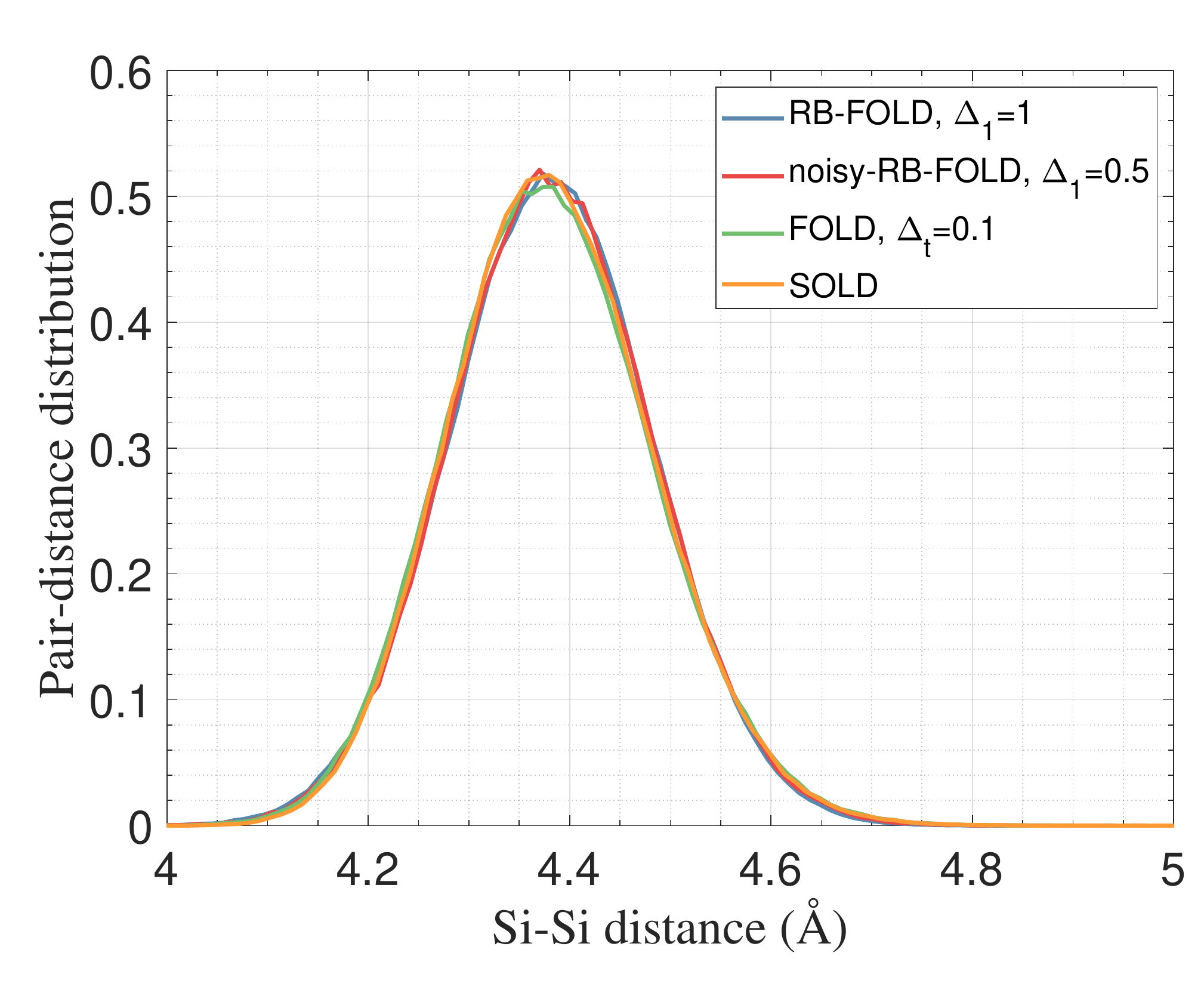}%
\end{minipage}%
\noindent\begin{minipage}[t]{1\columnwidth}%
\includegraphics[width=0.95\columnwidth]{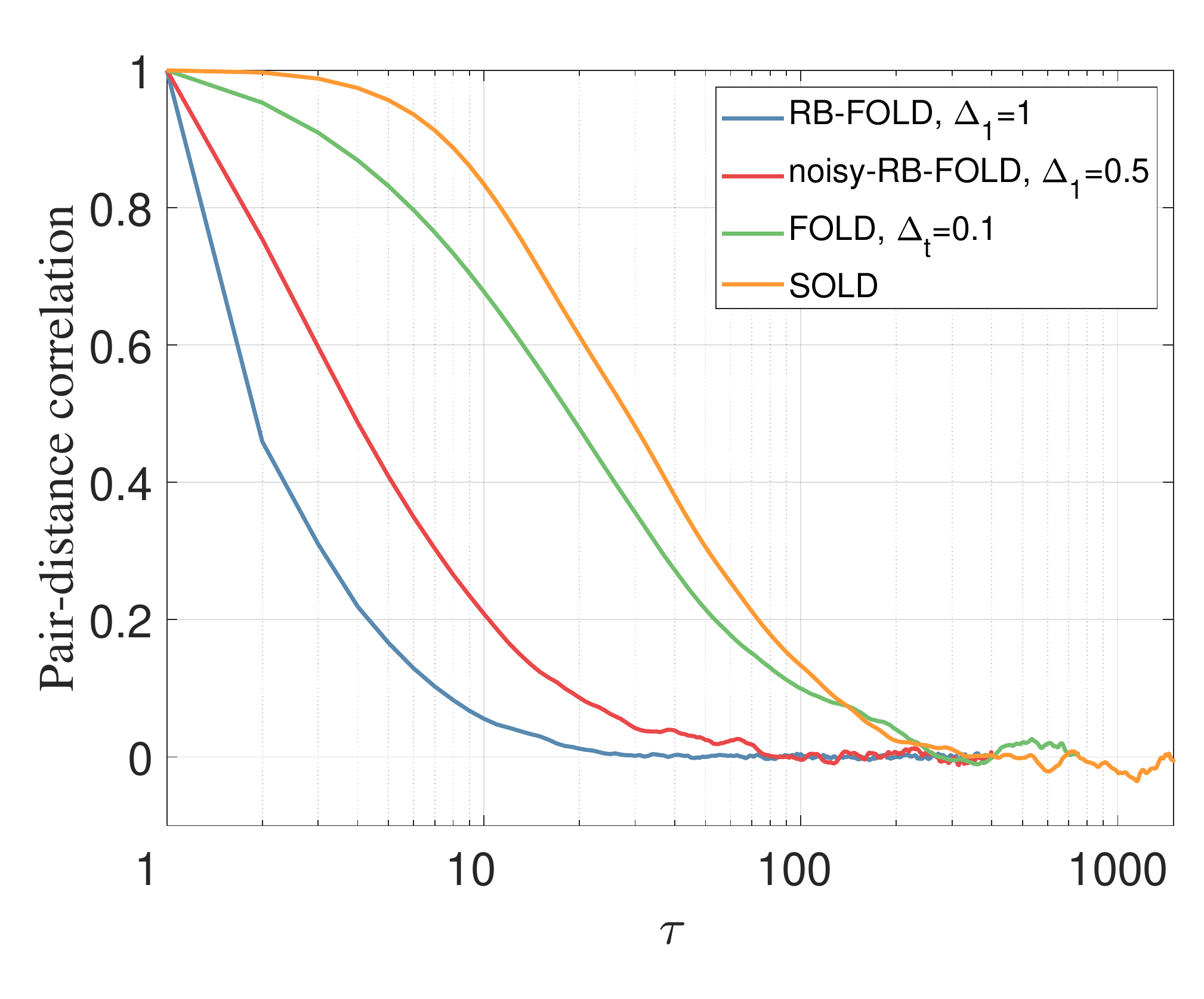}%
\end{minipage}\caption{\label{fig:Si35 validation}Comparison of four sampling methods, applied
to $\text{Si}_{35}\text{H}_{36}$ at $T=300K.$ The methods are: noisy-RB-FOLD,
based on sDFT and FOLD, RB-FOLD, and SOLD, based on dDFT. All FOLD-based
methods use optimal preconditioning $S=H$ (see supplementary material
concerning calculation of the Hessian $H$). For each of these methods
we produced a 3000-step trajectory starting from the same configuration.
From these four trajectories we average the Si-Si pair-distance distribution
function $g\left(r\right)$ (shown on the left panel) and the pair-distance
correlation functions $C_{\tau}$ (see supplementary material for
definition) in the right panel.}
\end{figure*}

We expect that noisy-RB-FOLD calculations to be highly efficient not
only for the Harmonic model but also for more realistic systems. To
demonstrate this, we apply the method to the problem of determining
the structural properties of a a realistic atomistic system such as
the $\text{Si}_{35}\text{H}_{36}$ nanocrystal at $T=300K$ described
with DFT at the the local density approximation level \footnote{All calculations in this work use real-space grids of spacing $\Delta x=0.5a_{0}$,
Troullier-Martins norm-conserving pseudopotentials~\citep{troullier1991efficient}
within the Kleinman-Bylander approximation~\citep{kleinman1982efficacious}.
Fast Fourier Transforms were used for applying the kinetic energy
operator and for determining the Hartree potentials and the method
of Ref.~\citenum{martyna1999areciprocal} was used for treating the
long range Coulomb interactions in a finite simulation cell with periodic
boundary conditions. DFT calculations were preformed under the local
density approximation (LDA) using the PW92 functional \citep{perdew1992accurate}.}. Our purpose is to validate the noisy-RB-FOLD sampling approaches
based on sDFT forces using calculations based on sampling methods
which employ dDFT forces (RB-FOLD, FOLD and SOLD \citep{arnon2017equilibrium})
and to compare the efficiencies of these methods. Note that all the
FOLD methods in the figure are based on optimal sampling, with $S=H$.
We could not show results for the choice $S=\alpha\text{cov}\boldsymbol{\phi}$
of Ref.~\citenum{mazzola2017accelerating} because of numerical problems
stemming from the fact that the sDFT forces have a force-covariance
matrix which is nearly singular (see supplementary material). In Fig.~\ref{fig:Si35 validation}
(left panel) we show that indeed our new noisy-RB-FOLD method as well
as the other methods predict the same first peak of the pair distribution
function $g\left(r\right)$ (to within statistical fluctuations) \footnote{$g\left(r\right)=\frac{\Delta_{n}(r)}{4\pi r^{2}\Delta_{r}\rho_{0}}$
where $\Delta_{n}(r)$ is the number of Silicon pairs at distance
$\left[r,r+\Delta_{r}\right]$ and $\rho_{0}$ average Silicon atom
density}. In order to study the efficiency, we plot in the right panel the
the pair-distance correlation function in term of the distance $r_{ij}$
between a pair of Silicon atoms, numbered $i$ and $j$:
\begin{equation}
C_{\tau}=\frac{\left\langle \sum_{t=1}^{N_{\tau}-\tau}r_{ij}^{t}r_{ij}^{t+\tau}\right\rangle _{\left\{ i,j\right\} }}{\left\langle \sum_{t=1}^{N_{\tau}-\tau}r_{ij}^{t}r_{ij}^{t}\right\rangle _{\left\{ i,j\right\} }}.
\end{equation}
where, $\left\langle \right\rangle _{\left\{ i,j\right\} }$ represents
an average over these pairs, and $N_{\tau}$ is the total number of
steps in the Langevin trajectory. $C_{\tau}$ has the initial value
of 1 at $\tau=0$ and then it decays non-monotonically as it settles
upon a steady fluctuation around zero. We define the time scale $\tau_{c}$
for this decay as the earliest time for which $C_{\tau_{c}}=0.1$.
Consider first the correlation functions for the FOLD and the SOLD
trajectories; both are seen to have a concave structure at small values
of $\tau$ which delays decay and turns convex only at much longer
times, and both trajectories exhibit a slow decay with $\tau_{c}\approx100$.
Next, consider the correlation functions for RB-FOLD: the deterministic
RB-FOLD with $\Delta_{1}=1$ ($\Delta_{t}=10$$,\Delta_{2}=0.5$)
and the noisy-RB-FOLD with $\Delta_{1}=0.5$ ($\Delta_{2}=0.375$)
(having an identical form as that calculated within deterministic
RB-FOLD with the same parameters, but not shown in the figure). In
the figure, $\tau_{c}$ is twice as large when $\Delta_{1}=0.5$ than
when $\Delta_{1}=1$, following our analysis above and both functions
have a similar convex form. We have verified that the correlations
of the noisy- and deterministic RB-FOLD trajectories for $\Delta_{1}=0.5$
are identical and we see that they represent an order of magnitude
improvement on the previously used SOLD approach for sDFT.

Summarizing, in previous work \citep{arnon2017equilibrium}, we used
SOLD to address the problem of noisy forces in sDFT calculations but
found that thousands of time steps were required to shake off the
correlations. Here, we developed a radically more efficient method
for sampling system configurations under stochastic forces. It capitalizes
on a recently proposed method \citep{mazzola2017accelerating} but
makes critical changes in the Langevin force sampling which restore
optimal preconditioning. The final procedure is to perform a random
walk following Eq.~(\ref{eq:RB-FOLD-STOCH}) while sampling the Langevin
forces from Eq.~(\ref{eq:RB-FOLD-zz-STOCH}).

Using a purely Harmonic model system we compared noisy-FOLD, and noisy-RB-FOLD
and showed that the latter is much more efficient and insensitive
to the time-step. We further showed that the noisy-RB-FOLD has similar
characteristics also when applied to real atomistic system using sDFT
forces. One notable difference between RB-DFT and noisy-RB-DFT concerns
with increasing the time step. One must assure that the left-hand
side of Eq.~\ref{eq:RB-FOLD-zz-STOCH} is positive definite, hence
at some point any increase of $\Delta_{1}^{2}/\Delta_{2}$ will necessitate
a reduction of $\text{cov}\boldsymbol{\phi}.$ This is especially
important at low temperatures. The results of this work provide a
general recipe for efficient and stable Boltzmann sampling under the
presence of stochastic forces. A good approximation to the Hessian
which is required here can perhaps be obtained from a force-field
calculation or from embedded fragment calculations used in the sDFT
procedure \citep{neuhauser2014communication,arnon2017equilibrium,fabian2019stochastic,chen2019overlapped}. 
\begin{acknowledgments}
RB gratefully acknowledges the support of the Bi-national US-Israel
Science Foundation under grant no. 2018368. D.N. acknowledges support
from the National Science Foundation, grant CHE-1763176. E.R. acknowledges
support from the Department of Energy, \emph{Photonics at Thermodynamic
Limits Energy Frontier Research Center, }under Grant No. DE-SC0019140.
\end{acknowledgments}

\bibliographystyle{unsrt}

\end{document}